\documentclass[showpacs,showkeys,PRB,twocolumn,aps,preprintnumbers,amsmath,amssymb]{revtex4-2}
\usepackage[breaklinks=true,colorlinks,citecolor=blue,linkcolor=blue,urlcolor=blue]{hyperref}
\usepackage{graphicx,dcolumn,bm,multirow,hyperref,mathtools,braket,color,url,epstopdf,amssymb,amsmath}
\usepackage{amsfonts}

\begin{document}


\title{Ultrafast Carrier Relaxation and Second Harmonic Generation in a Higher-Fold Weyl Fermionic System PtAl}

\author{Vikas Saini}
\email{vikas.saini@tifr.res.in}
\affiliation{Tata Institute of Fundamental Research, Mumbai 400005, India}

\author{Ajinkya Punjal}
\affiliation{Tata Institute of Fundamental Research, Mumbai 400005, India}

\author{Utkarsh Kumar Pandey}
\affiliation{Tata Institute of Fundamental Research, Mumbai 400005, India}

\author{Ruturaj Vikrant Puranik}
\affiliation{Tata Institute of Fundamental Research, Mumbai 400005, India}

\author{Vikash Sharma}
\affiliation{Tata Institute of Fundamental Research, Mumbai 400005, India}

\author{Vivek Dwij}
\affiliation{Tata Institute of Fundamental Research, Mumbai 400005, India}

\author{Kirtika Vijay}
\affiliation{Accelerator Physics and Synchrotrons Utilization Division, Raja Ramanna Centre for Advanced Technology, Indore, 452013, India \\ Homi Bhabha National Institute, Training School Complex, Anushakti Nagar, Mumbai, 400094, India}

\author{Ruta Kulkarni}
\affiliation{Tata Institute of Fundamental Research, Mumbai 400005, India}

\author{Soma Banik}
\affiliation{Accelerator Physics and Synchrotrons Utilization Division, Raja Ramanna Centre for Advanced Technology, Indore, 452013, India \\ Homi Bhabha National Institute, Training School Complex, Anushakti Nagar, Mumbai, 400094, India}

\author{Aditya Dharmadhikari}
\affiliation{Tata Institute of Fundamental Research, Mumbai 400005, India}

\author{Bahadur Singh}
\affiliation{Tata Institute of Fundamental Research, Mumbai 400005, India}

\author{Shriganesh Prabhu}
\email{prabhu@tifr.res.in}
\affiliation{Tata Institute of Fundamental Research, Mumbai 400005, India}

\author{A. Thamizhavel}
\email{thamizh@tifr.res.in} 
\affiliation{Tata Institute of Fundamental Research, Mumbai 400005, India}

\date{\today}


\begin{abstract}
In topological materials, shielding of bulk and surface states by crystalline symmetries has provided hitherto unknown access to electronic states in condensed matter physics. Interestingly, photo-excited carriers relax on an ultrafast timescale, demonstrating large transient mobility that could be harnessed for the development of ultrafast optoelectronic devices. In addition, these devices are much more effective than topologically trivial systems because topological states are resilient to the corresponding symmetry-invariant perturbations. By using optical pump probe measurements, we systematically describe the relaxation dynamics of a topologically nontrivial chiral single crystal, PtAl.   Based on the experimental data on transient reflectivity and electronic structures, it has been found that the carrier relaxation process involves both acoustic and optical phonons with oscillation frequencies of 0.06 and 2.94 THz, respectively, in picosecond time scale.  PtAl with a space group of $P$$2_{1}$3 allows only one non-zero susceptibility element i.e. $d_{14}$, in second harmonic generation (SHG) with a large value of 468(1) pm/V, which is significantly higher than that observed in standard GaAs(111) and ZnTe(110) crystals. The intensity dependence of the SHG signal in PtAl reveals a non-perturbative origin. The present study on PtAl provides deeper insight into topological states which will be useful for ultrafast optoelectronic devices.  
      
\end{abstract}

\maketitle

\section{INTRODUCTION}
Topological materials are enormously sought after, in condensed matter research, for their novel symmetry protected electronic structures and unique responses to electromagnetic radiations, which lead to numerous advanced potential applications ~\cite{zhang2017room,kovalev2020non,Ramamurthy,xu2020optical, ullah2020third,suo2022ultrafast,jadidi2020nonlinear,de2017quantized, sanchez2019linear}.
Topological phases are distinguished by various parameters, such as the degeneracies of the valence and conduction bands, the dimensionality of the crossing point, the chiral charge, and the tilting directions of the bands.   
These topological phases are the results of symmetry constraints present in the crystalline system combined with the leading interactions and spin-orbit coupling (SOC). Depending on symmorphic and non-symmorphic crystalline symmetries, we can distinguish two categories for the generic and time reversal invariant point band crossings, respectively. 
The non-symmorphic symmetry protection in electronic structure leads to a higher-fold band crossing at high symmetry points and exotic topology of bands.
This enables the observation of novel emergent quasi-particles in crystalline systems that are not observed as fundamental particles~\cite{yang2018symmetry,wang2016hourglass, wu2022nonsymmorphic}.

Apart from their electrical transport properties, topological materials are excellent platforms for studying the exotic physics of nonlinear optical responses and possess application potential due to their novel electronic structure and symmetry protections. Recent pump probe experiments on topological insulators such as Bi$_{2}$(Se, Te)$_{3}$ have shown bulk and surface state relaxations under the influence of terahertz radiation, making them good terahertz emitters~\cite{braun2016ultrafast, zhao2020generation}. In semimetals, the ultrafast carrier dynamics and unconventional SHG of the topological Dirac semimetal Cd$_{3}$As$_{2}$ have also been investigated, and distinctly Weyl semimetal TaAs manifests interesting optically controlled chiral anomalies~\cite{hajlaoui2012ultrafast, kovalev2020non,jadidi2020nonlinear, levy2020optical, Jadidi:17, kumar2011spatially, luo2013ultrafast}.

To be specific, PtAl crystallizes in the non-symmorphic space group $P2_{1}3$ that has a chiral crystal structure, similar to other chiral systems such as RhSi, CoSi, and PdGa etc. The first-principles calculations including SOC and symmetry analysis of this system unfold the existence of spin-3/2 exotic emergent Rarita-Schwinger (RS) fermions in the vicinity of the center ($\Gamma$ point) of the Brillouin zone (BZ). The expected dispersion relations of these quasiparticles have been experimentally observed recently~\cite{schroter2019chiral}. Likewise, the corner point $R$ of the cubic BZ exhibits six-fold degenerate band crossings, followed by two time-reversal copies of spin-1 excitations. The chiral charges of these exotic Weyl points at the $\Gamma$ and $R$ points are -4 and +4, respectively, which is significant and four times higher compared to other Weyl semimetals such as Ta(As, P) and Mo(Te, Se)$_2$~\cite{rees2020helicity,chang2017unconventional,sessi2020handedness,yao2020observation,li2019chiral,gao2019topological,chang2018topological, vikas,flicker2018chiral,vikas,wilde2021symmetry}.    

The predicted quantization value of the circular photogalvanic effect (CPGE) for PtAl is noteworthy, being four times higher than that of achiral Weyl semimetals due to the higher chiral charge (4) of its quasiparticles~\cite{de2017quantized,ni2020linear, chang2017unconventional, le2020ab}. This is crucial for the development of optoelectronic devices and sensors. The exotic bulk topological states and chirality of the crystal structure in PtAl leave curiosity and motivate the exploration of novel electromagnetic responses. 

We comprehend the ultrafast carrier relaxation mechanism of nontrivial quasiparticles (RS Weyl fermions) through optical pump probe measurements.
The observation of susceptibility tensor $d_{14}$ in strong second harmonic generation unequivocally proves the structural chirality in PtAl.

\section{METHODS}
{\it Single crystal growth--}
The single crystals of PtAl were prepared using the Czochralski method and the phase purity of PtAl is confirmed by the powder x-ray diffraction analysis, full details are given in reference~\cite{vikas}. 

{\it XPS and RPES experiments --}
Synchrotron X-ray photoemission spectroscopy (SR-XPS) and RPES measurements were performed at the undulator based Angle Resolved Photoelectron Spectroscopy beamline (ARPES BL-10), Indus-2. Atomically clean surface of the single crystal with similar bulk composition was obtained by argon ion sputtering at 1.5 keV for 30 min. SR-XPS core levels were recorded at h$\nu$= 700~eV with 0.3~eV energy resolution. RPES measurements carried out by recording the valence band (VB) spectra across the Pt $5p-5d$, Pt $4f-5d$ and Al $2p-3s$ transitions at the photon excitation energies ranging from 41~eV to 88~eV with energy resolution of 20~meV to 30~meV. All the photoemission measurements were carried out at low temperature of $\sim$20~K using SPECS Phoibos 150 electron energy analyzer. The base vacuum during the measurement was $\sim7~\times 10^{-11}$ mbar.

{\it Optical pump probe measurement--}
For optical pump probe measurement, we used an amplifier laser (Solstice Ace, 800~nm, 5.5~mJ, 1~kHz, 35~fs) which is split into two parts. One of the parts goes to Optical Parametric Amplifier (OPA) to generate 650~nm which is used to pump the sample and another part of the amplifier is used to probe the dynamics. The pump beam is focused on the sample using a 300~mm lens (Thorlabs LA1484) whereas the probe beam is focused using a 200~mm (Thorlabs LA1708-B) lens. The spot sizes of these pulses were measured using the beam profiler (Newport LBP2). The state of the polarization of the wavelengths generated by the amplifier and OPA is linear. To probe the dynamics in the system, the probe beam is scanned using a delay stage (Newport UTS100PP) at various temporal delays with reference to the pump beam. We used a balanced-photodetection (BPD) scheme and the signal is fed to a lock-in amplifier (SR830) with a modulated pump beam (Thorlabs MC2000B) as reference.

{\it Second harmonic generation--}
In the rotation anisotropy second harmonic generation  (SHG) polarimetry experiment, we used Ti:Sapphire amplifier laser (Solstice Ace, 1~kHz, 35~fs) with a central wavelength of 800~nm. A clean-up polarizer (CP) (Newport 10P109AR.16) was used to enhance the linear state of polarisation of the amplifier beam. The beam was modulated using a chopper (Thorlabs MC2000B-EC) for lock-in detection. The incident linear polarisation was rotated using a Quarter Wave Plate (QWP) (Newport 10RP54-2B) with measured ellipticity of ~0.98 and a linear polarizer (LP) (Newport 10P109AR.16). A 400~mm lens (Newport) was used to focus the light onto the sample. The angle of reflection and incidence is less than 3$^{\circ}$. The reflected fundamental and second harmonic (SH) frequency was picked by a mirror and directed through a series of spectral filters (EO Schott glass BG40, Thorlabs FESH0450) with a theoretical contrast ratio greater than 10$^{12}$ and a polarizer to an analyzer (Newport 10GT04) the second harmonic signal. The SH light was focused using a 50~mm lens (Thorlabs LA1213) onto the photodiodes (Hamamatsu S1223). The signal from the photodiode was amplified using a high-gain home-built current preamp and detected using a lock-in amplifier (SR830). The polarizer and analyzer were mounted on rotation stages (Newport PR50PP) to avoid angle errors. The power was measured using an integrating sphere (Ophir) and the spectra were measured using an Ocean Optics (FX series) spectrometer. 

{\it First principles calculations --}
Electronic structure calculations were performed within the density functional theory (DFT) framework using the Quantum Espresso package~\cite{giannozzi2009quantum, giannozzi2020quantum}. A dense 21$\times$21$\times$21 $k$-mesh was used for self-consistent calculations, using an energy cut-off of 60~Ry. Optical properties were calculated using the full potential linearized augmented plane-wave (FL-LAPW) method as implemented in WIEN2k package~\cite{schwarz2001augmented,schwarz2003solid,schwarz2003dft}. A $k$-mesh of 46$\times$46$\times$46 was used to obtain well-converged results.  


\section{RESULTS AND DISCUSSIONS}

\subsection{X-ray and Resonant Photoemission Spectroscopy}    
   

The PtAl single crystal has been grown by the Czochralski method~\cite{vikas}. The single crystal was characterized using X-ray and resonant photoemission spectroscopy. The details are given in the supplementary material (Figure~S1).

\begin{figure*}[]
	\includegraphics[width=1.0\textwidth]{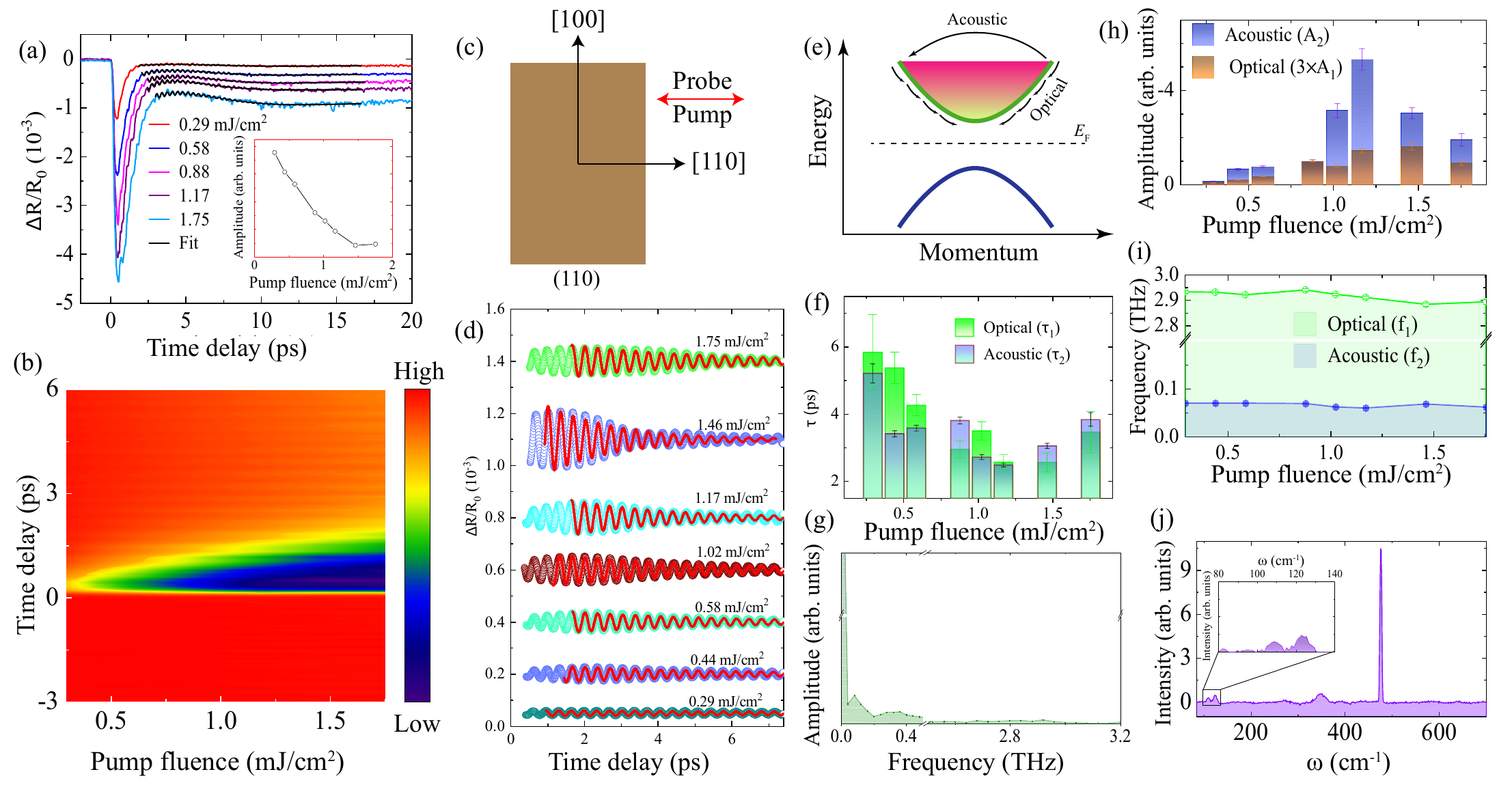}
	\caption{Pump probe experiments were performed with both linearly polarized pulses along the [110] direction. (a) The transient reflectivity is measured as a function of probe delay for various pump fluences up to 1.75~mJ/cm$^{2}$. Two exponential fits with oscillating terms are used to model the transient differential reflected signal, corresponding to oscillations frequencies of approximately 0.06 and 2.94~THz after thermalization. (b) A contour plot of the transient reflectivity ($\Delta R/R_{0}$) against various pump fluences for probe time delays of up to 6 picoseconds. (c) A schematic of the pump and probe polarization directions on the sample surface. (d) The background subtracted optical phonon oscillations of the transient reflectivity correspond to various pump fluences. (e) Schematic illustration of carrier relaxation in the conduction band using optical and acoustic phonons. (f) The phonon lifetimes ($\tau_{1}$, $\tau_{2}$) were estimated for both decay channels and plotted against the pump fluence. (g) Fourier transform of $\Delta R/R_{0}$ against the time delay at a pump fluence of 1.17~mJ/cm$^{2}$. (h) The proportionality weight coefficients (A$_{1}$, A$_{2}$) of the phonon relaxations are plotted against the pump fluence. (i) The phonon oscillation frequencies are extracted from the overlapped oscillating $\Delta R/R_{0}$ signal and plotted against various pump fluences. (j) Experimentally observed Raman spectrum for PtAl at room temperature.}\label{Fig1}
\end{figure*}

\subsection{Optical Pump Probe Spectroscopy}
The carrier relaxation mechanism of the excited quasiparticles in PtAl is investigated by the ultrafast pump probe technique.
The pump and probe pulses of a repetition rate of 1~kHz and energies of 1.9 eV and 1.55 eV, respectively, are incident on the (110)-plane and are polarized along the same direction, as depicted in Figure~\ref{Fig1}(c).
In addition, the pump and probe pulses are incident at the same position on the sample surface. The probe pulse coincides with the pump pulse for uniform excitations, where the respective diameters are 350 and 660~$\mu$m.
Initially, the pump pulse injects charge carriers from the valence band (VB) to the conduction band (CB) and the relaxation mechanism of these excited carriers is tracked using a probe pulse, which is incident on the sample after a short delay following the pump pulse.

Since PtAl is a semimetallic system with a relatively large density of states, the reflection of carriers dominates over transmission in contrast to semiconductors for infrared (IR) radiation.
Experimentally the signal $\Delta R$/$R_{0}$ $=$ ($R-R_{0}$)/$R_{0}$ is measured; where R and $R_{0}$ are the transient reflections with and without of the pump pulse, respectively~\cite{kumar2011spatially}.
Carrier relaxation dynamics of PtAl is discussed in Figure~\ref{Fig1}.
In negative time delays, the probe pulse arrives before the pump pulse on the sample surface and is completely reflected, resulting in zero differential reflectivity as observed in Figure~\ref{Fig1}(a).
When the time delay becomes positive, the probe pulse is absorbed by the system leading to the sharp negative dip of $\Delta R/R_{0}$ in the transient reflectivity. Beyond 0.4~ps, the system tries to recover back to the equilibrium state therefore the magnitude of $\Delta$R/$R_{0}$ decreases towards zero as time goes by.

\begin{figure*}[]
	\includegraphics[width=1.0\textwidth]{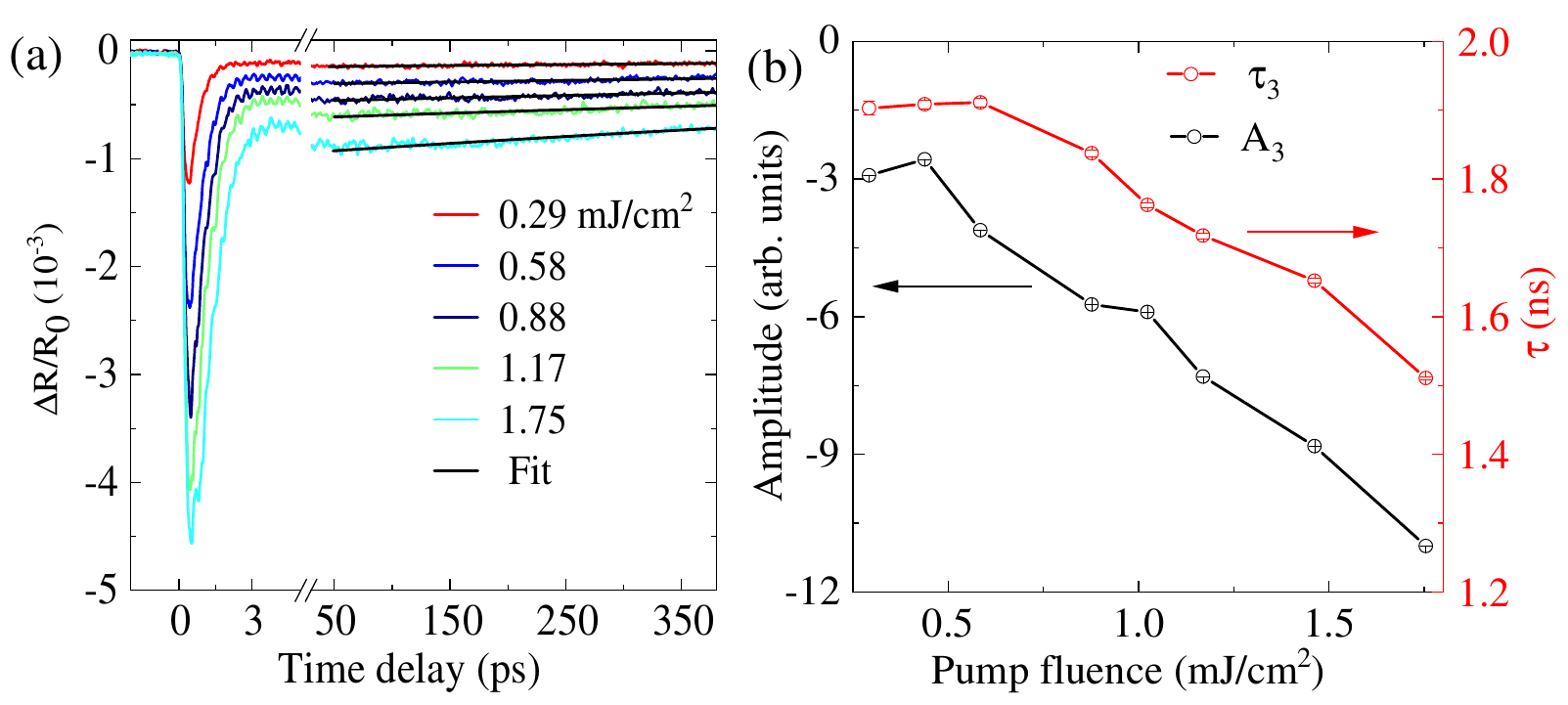}
	\caption{(a) $\Delta$R/R$_{0}$ is plotted in the higher probe delay regime for slow relaxation, where the black fitted line corresponds to a single exponential term for carrier recombination. (b) The parameters $A_{3}$ and $\tau_{3}$ for electron-hole recombination as a function of incident pump fluence.}
	\label{Fig2}
\end{figure*}

The $\Delta$R/$R_{0}$ with various pump energies is depicted in the Figures~\ref{Fig1}(a) and ~\ref{Fig2}(a). With increasing pump fluence the maxima of the $\Delta$R/$R_{0}$ increases linearly which signify the carriers get injected into the conduction band by a single photon absorption of the pump energy nearly up to 1.17~mJ/cm$^{2}$~\cite{zhang2017transient, cui2014transient}. The linearity is shown in the inset of Figure~\ref{Fig1}(a).
As the pump fluence increases, more carriers are injected into the conduction band, leading to an increase in the amplitude of the transient reflectivity dip. This amplitude is related to the temperature and density of hot electrons. Specifically, the electron temperature is highest during the carrier injection phase, after which the hot electrons gradually cool down, as this is observed from the color gradient in Figure~\ref{Fig1}(b).
After carriers are injected into the conduction band, the excited hot carriers try to come down to the VB by lowering their energy through various mechanisms, such as electron-electron, electron-phonon, phonon-phonon scatterings, and electron-hole recombinations~\cite{allen1987theory, ULSTRUP2015340, breusing2011ultrafast}. The electron-electron intraband scatterings are faster than the electron-phonon and phonon-phonon scatterings which appear in the time scale of 10 to a few hundred femtoseconds~\cite{george2008ultrafast}.
Furthermore, electron-phonon scatterings occur on the picosecond timescale, whereas interband electron-hole recombinations typically occur on a much slower timescale in the range of 100 to 1000~ps~\cite{taneda2007time,onishi2015ultrafast}.

The electron temperature for a particular pump fluence is estimated using the following expression~\cite{luo2012quasiparticle, liu2021dirac,wu2020ultrafast}
\begin{equation}
	T_e = \left\langle\sqrt{T_l^2 + \frac{2 (1-R) F}{\gamma_v \delta} \exp{(-z/\delta)}}\right\rangle
\end{equation}
Where $T_l$ is the lattice temperature, in the present case the pump probe measurements have been carried out at room temperature and hence it is considered as 300~K. $R$ is the reflectivity of the pump pulse, which we calculated from first-principles calculations as $\sim$0.535 at 650~nm. $F$ denotes the pump fluence. The skin depth ($\delta$) is also calculated and found to be approximately 16.6~nm at 650~nm. $\gamma_V = \gamma \rho$, where $\gamma$ is the Sommerfeld coefficient which can be estimated from the heat capacity measurements. From the low-temperature heat capacity data we have estimated $\gamma$ as $\sim$0.86 mJ/mole-K$^{2}$ (see Figure~S2 in the suppl. material). The $\rho$ represents the molar density $\frac{1}{N_{A} V}$; $N_{A}$ represents the Avogadro's constant, and $V$ is the unit cell volume. For PtAl, the lattice constant $a$ = 4.864$\AA$, and hence the unit cell volume is nearly~115$\AA$$^{3}$. Here, $z$ represents the distance from the sample surface, and that is taken $\delta/2$ as an average of the skin depth.
The $T_e$ increases as the pump fluence increases, at 0.29~mJ/cm$^{2}$ it is estimated to be $2848$~K that increase up to $6944$~K for the fluence 1.75~mJ/cm$^{2}$.

The oscillating transient reflectivity signal measured for PtAl against probe delay exhibits two frequencies that overlap. Transient reflectivity oscillations can arise for a variety of reasons, such as surface plasmons, coherent optical phonons, strain waves, interference between the reflected probe beams from the top and bottom layers of the sample, and coherent acoustic phonons. These oscillations have been observed for systems ranging from thin films to bulk materials~\cite{kumar2011spatially, Thomsen, Thomsen2,tachizaki2006scanning}.
For PtAl, the optical phonon frequency is nearly unchanged when the spot positions of the pump and probe pulses are changed together on the sample surface, which eliminates the possibility of oscillations caused by surface plasmons. The existence of optical phonons in PtAl has been verified from the experimentally observed Raman spectrum (Figure~\ref{Fig1}(j)), which shows an optical Raman peak around 3.2~THz, consistent with the higher frequency obtained in the oscillations (Figure~\ref{Fig1}(i)).
The background subtracted oscillations corresponding to the coherent optical phonons are shown in Figure~\ref{Fig1}(d) at various pump fluences, along with the fitted corresponding oscillating term in Eq.~\ref{Eq1}.
Furthermore, we have calculated the skin depth ($\delta$) using first-principles calculations ($\delta$ $=$ $1/\alpha$), where $\alpha$ is the absorption coefficient and is found to be 0.015 $\mu$m for $\lambda$ $\sim$ 800~nm. This is much smaller than the sample thickness ($\sim$50 $\mu$m), so we can safely conclude that the strain waves and interference of reflected beams are not considered. Therefore, it is likely that the small frequency in the transient reflectivity is a consequence of the coherent acoustic phonon oscillations, as observed for Bi$_{2}$Se$_{3}$~\cite{kumar2011spatially}.
\begin{figure}[]
	\begin{center}	\includegraphics[width=0.5\textwidth]{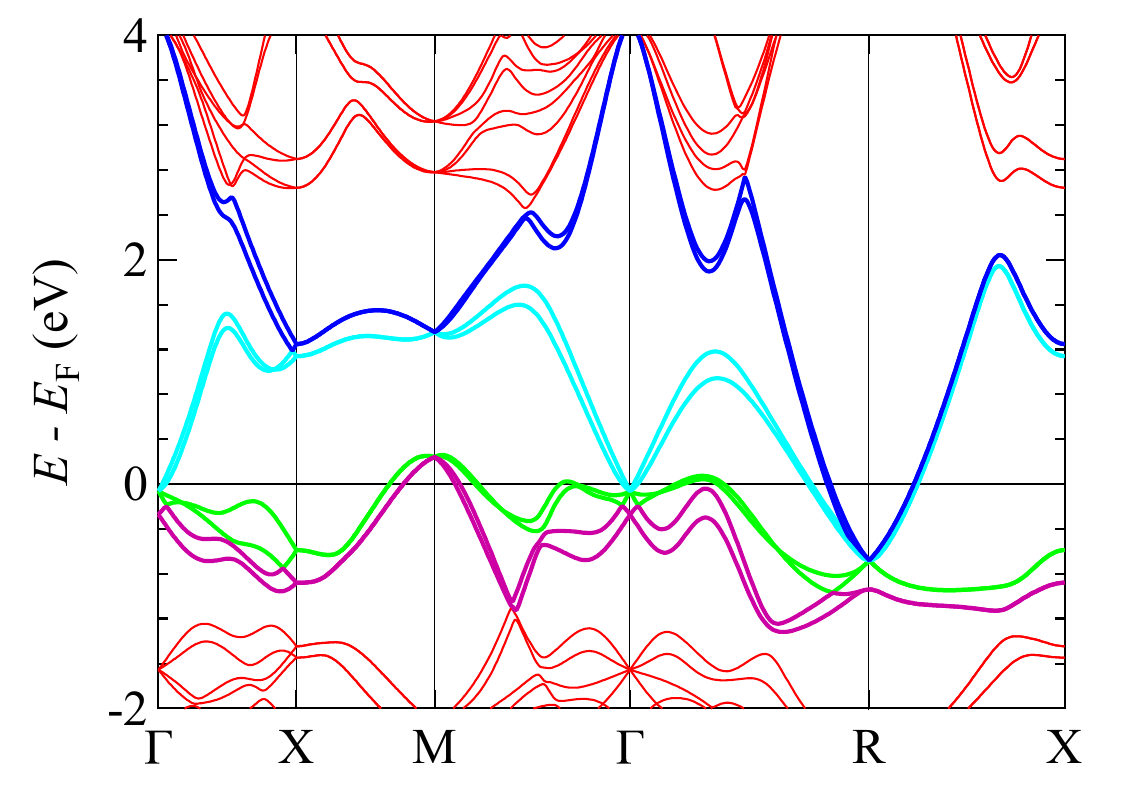}
		\caption{ Bulk band structure of PtAl calculated using the GGA exchange correlation functional in presence of SOC. Four pink and green color bands are the valence bands whereas the other four bands crossed at the Fermi level shown in the cyan and blue colors represent the conduction bands.}
		\label{Fig3}
	\end{center}
\end{figure}

In PtAl, pump excitations take place in the 0.4~ps for the 0.29~mJ/cm$^{2}$ pump fluence and further, it increases slightly with higher pump energies as observed from Figure~\ref{Fig1}(a) and consequently thermalization from the coulomb scatterings appear approximately in 0.6~ps where the transient reflectivity ($-\Delta R/R_{0}$) decreases with the increasing time delay followed by less absorption of the probe energy of the excited carriers. 
After the process of thermalization, we observed that $-\Delta R/R_{0}$ does not decrease monotonically towards zero rather it increases and later it starts to decrease as observed in Figure~\ref{Fig2}(a). This behavior of the transient reflectivity in PtAl is understood from the overlap of acoustic phonon oscillations carrying 0.06~THz frequency with another optical phonon frequency 2.94~THz on the transient reflectivity, which is observed for approximately up to 15~ps time delay (Figure~\ref{Fig1}(a)). The Fourier transform of transient reflectivity depicts the oscillation frequencies as shown in Figure~\ref{Fig1}(g) in agreement with the observed oscillation frequencies obtained using two exponent fitting (see Figure~\ref{Fig1}(i)).
The carrier relaxation process in PtAl is illustrated in Figure~\ref{Fig1}(e). After thermalization, hot electrons lose their energy through both optical and acoustic phonons until they rapidly reach the bottom of the conduction band~\cite{wu2020pump}.
After approximately 2~ps, the relaxation is understood by the three exponent relaxation terms as described below~\cite{ye2018nonlinear}

\begin{equation}
	\normalsize
	\label{Eq1}
 \begin{split}
	\frac{\Delta R}{R_{0}} = A_{1}exp{(-t/\tau_{1})} sin(2\pi f_{1} t + \phi_{1}) + \\
	 A_{2} exp{(-t/\tau_{2})} sin(2\pi f_{2} t + \phi_{2}) + A_{3} exp{(-t/\tau_{3})}
	\end{split}
	\normalsize
\end{equation}                                         
Here $A_{1}$, $A_{2}$, and $A_{3}$ are the coefficients associated with the respective decay channels, and $f_{1}$ and $f_{2}$ are the related frequencies of optical and acoustic phonons. $\tau_{1}$, $\tau_{2}$, and $\tau_{3}$ are the phonon lifetime for their respective decay channels. From the above equation 2.94~THz electron-phonon relaxation is associated with the first term $A_{1}$ and the other 0.06~THz relaxation is associated with the second term $A_{2}$. Notably, the third term of the slow relaxation appears in large time delay as fitted beyond 50~ps which is attributed to the recombination of hole and electron carriers. The first two-terms fitted transient reflectivity signal is depicted in Figure~\ref{Fig1}(a) for various pump fluences.

From the three exponents fitting we extracted the above mentioned parameters which explain the process of carrier relaxation in PtAl. The obtained parameters $A_{1}$ and $A_{2}$ from the fitting are plotted against pump fluences and shown in Figure~\ref{Fig1}(b). 
The coefficients obtained from the optical and acoustic phonon relaxation processes are primarily dominant at small time delays between the pump and probe beams, typically up to 15~ps after thermalization. The decay of acoustic phonon is prominent for all pump fluences compared to the optical phonon as shown in Figure~\ref{Fig1}(h). More pump energy implies more carriers are being injected into the CB. The optical relaxation coefficient $|A_{1}|$ increases gradually as the pump fluence is increased. On the other hand, the coefficient $|A_{2}|$ for acoustic phonon relaxation initially increases with increasing pump fluence up to 1.17~mJ/cm$^{2}$. However, beyond 1.17~mJ/cm$^{2}$, it decreases due to the increasing number of excited carriers in the CB that suppress relaxation through this channel as more carriers are injected into the CB. 
The lifetime of acoustic phonons is roughly of the same order as that of optical phonons, on the order of a few picoseconds as depicted in Figure~\ref{Fig1}(f), that is higher than that of in the Bi$_{2}$Se$_{3}$~\cite{kumar2011spatially} and monolayer MoS$_{2}$~\cite{seo2016ultrafast}. The decrease in the relaxation time against pump fluence is also observed in other systems like $\alpha$-In$_{2}$Se$_{3}$ nanoflake, WS$_{2}$ monolayer, GaN/InGaN quantum well nanowires etc.~\cite{wang2019layer, kuroda2020dark, wu2020pump,wolfson2018long,boubanga2016ultrafast}
Both acoustic and optical phonon lifetimes decrease with increasing pump fluence up to 1.17~mJ/cm$^{2}$, indicating an increase in the strength of electron-phonon coupling ($\lambda$), as the relation with relaxation rate is discussed in the Refs.~\cite{wu2020ultrafast, brorson1990femtosecond}. The phonon frequencies of related phonon oscillations remain nearly unchanged with increasing pump fluence as shown in Figure~\ref{Fig1}(i).

For the slow relaxation, we have fitted a single exponential function for probe delays above 50~ps, as shown in Figure~\ref{Fig2}(a). Relaxation time at the 0.58~mJ/cm$^{2}$ pump fluence is around 1.9~ns which further decays with an increase in the pump fluence resulting in relatively faster electron-hole recombination~\cite{gupta1992ultrafast, PhysRevB.92.161104}. Alongside the proportionality coefficient $A_{3}$ also increases in magnitude with increasing pump fluence, which reveals that more carriers are recombined at higher pump energy as shown in  Figure~\ref{Fig2}(b).   

To understand the process of equilibrium of the injected carriers we look into the electronic structure of PtAl. 
However, there are electronic bands along all high symmetry paths near the Fermi level, therefore for the unpolarized pump pulse all directional excitations are possible. Here, pump and probe polarizations are along [110] crystallographic direction, which is equivalent to the $\Gamma$ - M direction in momentum space. Thus, only the optical excitations of this direction are allowed for the valence electrons of mostly green bands in Figure~\ref{Fig3} by the 1.9 eV pump pulse to the conduction band of the near M point, as shown in the blue colors in Figure~\ref{Fig3}.
After excitations, hot carriers distribute themselves based on the Fermi-Dirac statistics to lower the electron temperature in the order of 0.6~ps time scale. This process of thermalization takes place in the CB denoted in the blue color. 
The valleys located at the X and M points are formed by the same blue color CB, which is broad in momentum space. As a result, excited carriers can be easily scattered towards the edges of these valleys of the M and X points (M $\leftrightarrow$ X) by emitting acoustic phonons of very low energy (see schematic in Figure~\ref{Fig1}(e)). This is consistent with our experimentally observed acoustic phonons of $\sim$0.06~THz.
The relaxations of the carriers via optical phonon emissions take place in the same probe delay regime where oscillations from acoustic phonons are observed as depicted in Figure~\ref{Fig1}(a). Later, once the excited carriers are at the bottom of the CB they recombine with the hole carriers present in the VB in the order of nanosecond timescale as discussed in Figure~\ref{Fig2}(b).
Interestingly, PtAl exhibits structural chirality due to the absence of rotation and inversion symmetry, which allows observation of second harmonic generation through inversion asymmetry. Additionally, PtAl electronic bands are rich in Berry curvature near the Fermi level. Therefore, we investigate the role of Berry curvature of the electronic states in the second harmonic generation.
\subsection{Second Harmonic Generations}
\begin{figure*}[]
	\includegraphics[width=1.0\textwidth]{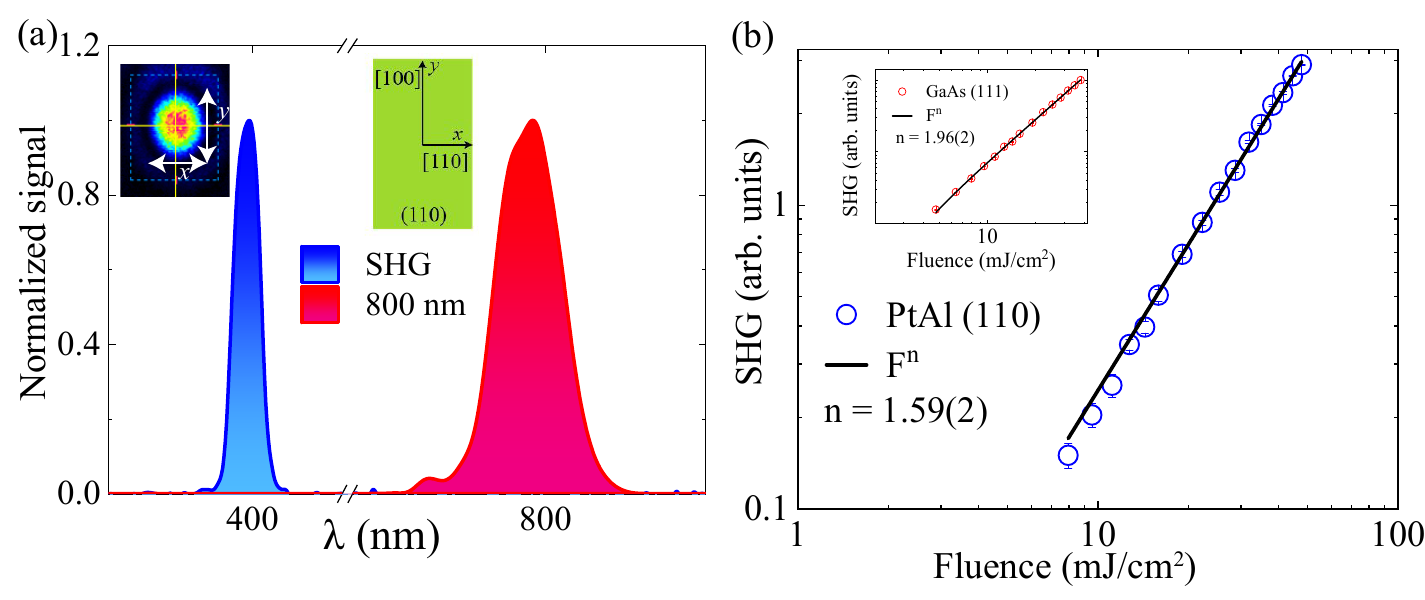}
	\caption{Second order nonlinear optical response of PtAl. (a) The 800~nm incident near IR radiation shows the second harmonic around 400~nm and the left inset is a spot size profile of the incident beam where $x$ 
		and $y$ dimensions are 184 and 200~$\mu$m, respectively. The right inset denotes the polarization axes in the lab frame with respect to the crystallographic directions. (b) The SHG power of PtAl is measured for several pump fluences, with the inset showing the SHG of GaAs for comparison. A fluence-dependent term $I$ $\propto$ $F^{n}$ is fitted, where $n$ is the fitting parameter.}
	\label{Fig4}
\end{figure*}

In this section, we will discuss the second order nonlinear optical response of PtAl in the near infrared regime, specifically at 800~nm.
Figure~\ref{Fig4}(a) depicts the SHG by an 800~nm laser pulse at room temperature that shows a peak around $\lambda$ = 400~nm. The incident beam profile is shown in the inset.
We have also performed SHG on a GaAs(111) sample, which is well known for exhibiting second order optical response. When comparing the SHG signals of PtAl with standard nonlinear materials such as GaAs and ZnTe, we observed that PtAl generates a stronger SHG signal, as discussed later in this section.     
The SHG signal of PtAl has been analyzed with different pump fluence, as shown in Figure~\ref{Fig4}(b) up to 50~mJ/cm$^{2}$. An increase in the fluence increases the SHG signal at 400~nm. It has been observed that for fluences above $\sim$ 25 mJ/cm$^{2}$, the relationship between the fluence and response deviates from the expected quadratic behavior, which is governed by semiclassical perturbation theory. The fitting exponent (n) for PtAl is 1.59(2). The SHG response was also measured for the standard GaAs(111) material as shown in the inset of Figure~\ref{Fig4}(b), the exponent for GaAs was obtained as 1.99(1), which is very close to the expected value of 2. Therefore, the deviation from the expected quadratic behavior has also been observed in other topologically nontrivial materials such as Cd$_{3}$As$_{2}$, Ta(As, P), RhSi which is attributed to the finite Berry curvature of the wavefunctions incorporating single and two-photon interband transitions, as shown by calculations for the RhSi compound~\cite{kovalev2020non,li2018second, wu2017giant, lu2022second}.

\begin{figure*}[]
	\includegraphics[width=1.0\textwidth]{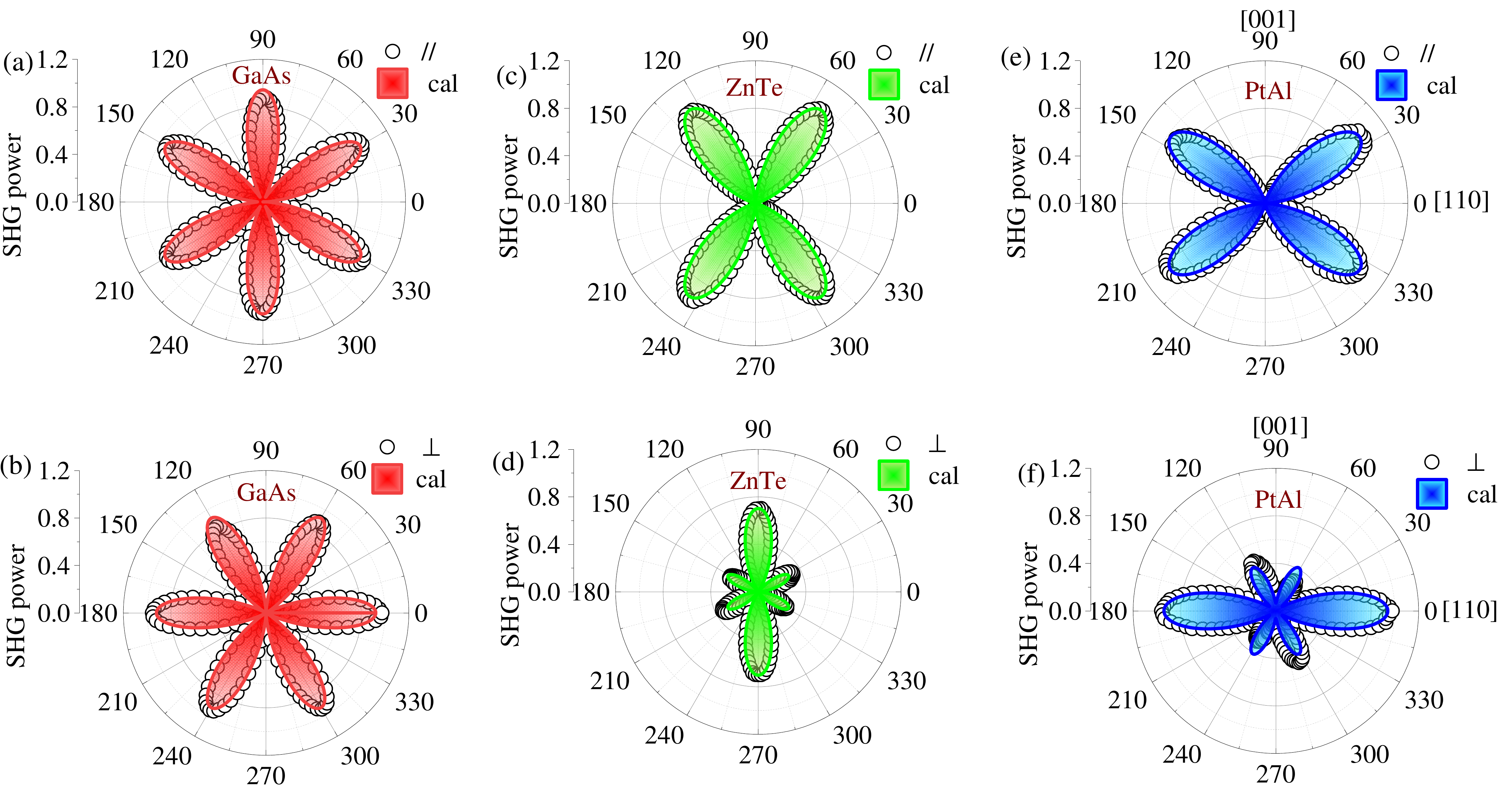}
	\caption{The SHG was measured for the materials GaAs(111), ZnTe(110), and PtAl(110). The normalized SHG power for GaAs is shown in (a) and (b), for ZnTe in (c) and (d), and for PtAl in (e) and (f). The measurements were taken for each parallel and perpendicular crossing of the polarizer and analyzer at nearly perpendicular incidence on the sample for an 800~nm wavelength.}
	\label{Fig5}
\end{figure*}  

As mentioned earlier, PtAl is a chiral crystal with a space group of $P2_{1}3$, which has two-fold screw rotation symmetry but lacks inversion, rotational, and mirror plane symmetries. The point group $23(T)$ for PtAl allows only one non-zero susceptibility element, $\chi_{xyz}$ $=$ $2d_{14}$. For the cubic crystal structure point group $T$ suggests the coexistence of a three fold and two-fold rotational symmetries thus the nonzero $d_{14}$ confirms the chiral crystal structure of PtAl. The polarization tensor for the second-order response in a nonlinear medium can be expressed as follows 

\begin{equation*}
	\begin{bmatrix}
		P(x) \\
		P(y) \\
		P(z) 
	\end{bmatrix}
	=\begin{bmatrix}
		d_{11} &  d_{12} & d_{13} & d_{14} & d_{15} & d_{16}\\
		d_{21} &  d_{22} & d_{23} & d_{24} & d_{25} & d_{26}\\
		d_{31} &  d_{32} & d_{33} & d_{34} & d_{35} & d_{36}
	\end{bmatrix}
	\begin{bmatrix}
		E_{x}^{2} \\
		E_{y}^{2} \\
		E_{z}^{2} \\
		2E_{y}E_{z} \\
		2E_{x}E_{z} \\
		2E_{x}E_{y} \\
	\end{bmatrix} 
\end{equation*}

Where E$_{x}$, E$_{y}$, and E$_{z}$ are the electric field components of the incident polarized light in the lab frame on the sample.         

The SHG signal of PtAl is analyzed using linearly polarized light incident normally on the (110) crystallographic plane of the sample as illustrated in the right inset of Figure~\ref{Fig4}(a), this also depicts the in-plane crystallographic directions. 
Before the beam is incident on the sample, it passes through a polarizer to make linear polarization of the incident beam. The reflected light then passes through an analyzer before being detected by a photodiode.
The incident light polarization makes an angle $\theta$, with the PtAl crystal axis [110], so the in-plane components of the electric field, $E_x$ and $E_y$, can be expressed as $E_x = E_0 \cos(\theta)$ and $E_y = E_0 \sin(\theta)$ along [110] and [100] crystallographic directions, where $E_0$ is the amplitude of the electric field and $E_z = 0$. The angular dependence of SHG has been measured for PtAl and for comparison we have also measured the angular dependence of the standard samples GaAs and ZnTe by rotating both the polarizer and analyzer simultaneously in two configurations of parallel and perpendicular crossings.

The transformed $d$ matrix from the crystal frame to the lab frame is as follows:

\begin{equation*}
	d =
	\begin{bmatrix}
		0 &  0 & 0 & 0 & 0 & -d_{14}\\
		-d_{14} &  0 & d_{14} & 0 & 0 & 0\\
		0 &  0 & 0 & d_{14} & 0 & 0
	\end{bmatrix}
\end{equation*}

The intensity components, $I_{\parallel}(\theta)$ and $I_{\perp}(\theta)$, are measured using the equations $I_{\parallel}(\theta) = |P(x) \cos(\theta) + P(y) \sin(\theta)|^2$ and $I_{\perp}(\theta) = |P(x) \sin(\theta) - P(y) \cos(\theta)|^2$, respectively.
To analyze the angle dependence of the SHG, the SHAARP.$si$ package is used~\cite{zu2022analytical}. The code requires the refractive index at 800 nm and 400 nm. These values, n($\omega$) = 4.58 + 3.91$i$ and n($2\omega$) = 2.89 + 2.71$i$, were obtained from the first principles calculations.

The polarization dependence of the SHG for GaAs(111), ZnTe(110), and PtAl(110) is analyzed for the parallel and perpendicular crossings as depicted in Figure~\ref{Fig5}. The obtained $d_{14}$ parameter for GaAs and ZnTe are $\sim$311(1) and $\sim$97.4(1)~pm/V, respectively which is in close agreement with the previously reported values~\cite{wagner1998dispersion, wu2017giant}.

The $d_{14}$ coefficients obtained for the parallel and perpendicular crossings of polarization in PtAl are around $\sim$468(1) and $\sim$516(5)~pm/V, respectively.
There are recent optical SHG measurements on the type-II Weyl semimetal $\beta$-WP$_{2}$, leading to a strong SHG signal in the bulk materials~\cite{hu2022strong}. 
The $d_{14}$ coefficient for PtAl is higher than those measured for standard GaAs(111), ZnTe(110), and $\beta$-WP$_{2}$ compounds, suggesting that PtAl is a potential candidate for SHG.     

\section*{CONCLUSIONS}
We prepared a single crystal of PtAl using the Czochralski method and carried out optical measurements along with XPS measurements. The pump probe experiment results showed the relaxation of the bulk topological states, including a slow electron-hole pair recombination. 
The SHG in PtAl does not follow a conventional square law dependence on incident intensity, as predicted by semiclassical perturbation theory due to interband transitions of carriers.
The obtained $d_{14}$ coefficient for PtAl is significantly high as compared to standard samples such as ZnTe and GaAs. The comprehensive analysis of the experimental data enhances our understanding of the PtAl system and could be useful for further studies and potential technological developments.     

\section*{Acknowledgements}      
We acknowledge Professor Jyotishman  Dasgupta and Debojyoti Roy for their help in Raman measurements. We acknowledge Professor V. Gopalan and Zu Rui for their expert help in SHAARP program. 
We also acknowledge TIFR central workshop, Atul Raut, Gajendra Mulay and Rodney Bernard for their support in building experimental set-ups.

\section*{Data Availability}
The data supporting the findings of the study is available and can be requested from the corresponding author.

%

\end{document}